\documentclass[12pt]{article}
\pdfoutput=1
\usepackage{jheppub}
\usepackage{simplewick}
\usepackage{verbatim}
\usepackage{graphics, color}
\usepackage{graphicx}
\usepackage{amsmath}
\usepackage{amssymb}
\usepackage{amsthm}
\usepackage{amsfonts}
\usepackage[usenames,dvipsnames]{xcolor}
\usepackage[normalem]{ulem}

\newcommand{\be}{\begin{equation}}
\newcommand{\ee}{\end{equation}}

\newcommand{\lan}{\langle}
\newcommand{\ran}{\rangle}

\newcommand{\mO}{\mathcal{O}}

\definecolor{grey}{rgb}{.5,.5,.5}
\definecolor{bluegreen}{rgb}{0,.5,.5}
\definecolor{darkgreen}{rgb}{0,.5,0}

\newcommand{\beq}{\begin{equation}}
\newcommand{\eeq}{\end{equation}}
\def\({\left(}
\def\){\right)}

\begin{document}

\title{Conformal Symmetry Breaking and Thermodynamics of Near-Extremal Black Holes}
\author{Ahmed Almheiri,}
\author{Byungwoo Kang}
\affiliation{Stanford Institute for Theoretical Physics, Department of Physics, Stanford University, Stanford, CA 94305, USA}
\emailAdd{almheiri@stanford.edu}
\emailAdd{bkang@stanford.edu}
\abstract{
It has been argued recently by Almheiri and Polchinski that the near-horizon conformal symmetry of extremal black holes must be broken due to gravitational backreaction at an IR scale linear in $G_N$. In this paper, we show that this scale coincides with the so-called `thermodynamic mass gap' of near-extremal black holes, a scale which signals the breakdown of their thermodynamic description. We also develop a method which extends the analysis of Almheiri and Polchinski to more complicated models with extremal throats by studying the bulk linearized quantum field theory. Moreover, we show how their original model correctly captures the universal physics of the near-horizon region of near-extremal black holes at tree level, and conclude that this equivalence of the conformal breaking and mass gap scale is general.
}
\maketitle
\section{Introduction}

Extremal black holes have taken center stage in the modern development of quantum gravity. They provided a concrete example where the Bekenstein-Hawking entropy was fully accounted for by microstate counting using perturbative string theory techniques \cite{Strominger:1996sh}, and furthermore had a central role in the advent of AdS/CFT \cite{Maldacena:1997re}. Despite these developments, many features of extremal black holes remain puzzling.

Perhaps the most famous one is their large zero temperature entropy. At zero temperature, the macroscopic horizon area of the black hole in Planck units presumably provides a count of the number of ground states, at fixed charge, of some quantum mechanical system. In the absence of supersymmetry, it is not clear what symmetry protects this huge degeneracy. We will not address this issue in this paper, but any progress on it would be exciting.

Another important puzzle about the extremal black holes is regarding their dynamical degrees of freedom. Take, for example, a spherical charged extremal black hole residing in $AdS_4$. Its geometry interpolates between $AdS_4$ in the UV and $AdS_2 \times S^2$ in the IR, representing how the dual boundary description is modified under RG flow. The $AdS_2 \times S^2$ description represents the IR fixed point of the boundary field theory. Therefore, it would be natural to interpret  the low energy excitations of the field theory as describing excitations propagating on the $AdS_2 \times S^2$ background. From the scaling symmetry of $AdS_2$, it can be shown that the spectrum of these excitations must attain the form \cite{Jensen:2011su}
\be  \label{density}
\rho(E) = A\delta (E)+B/E,
\ee       
for some dimensionless constants $A$ and $B$. The second term in this expression is problematic; it predicts a continuous spectrum as well as an infinite number of states below any given energy inconsistent with the boundary theory being defined on a finite volume. One generically expects the finite volume to induce a discrete spectrum. Setting $B=0$, the entire IR spectrum of the theory is described by the ground state degeneracy. This conclusion is also problematic as it would preclude all dynamics in the theory; all correlation functions would be time independent. This is in direct tension with the bulk expectation that long time behavior of correlators in this background adopt the conformal form in time.

This conclusion about the spectrum was also arrived at in \cite{Maldacena:1998uz} who showed that it was not possible to maintain the $AdS_2$ asymptotics for finite-energy states. This was further elaborated on in \cite{Almheiri:2014cka} in a two-dimensional dilaton gravity toy model chosen to exhibit generic behavior of spacetimes whose IR geometry is $AdS_2 \times X$ for some compact space $X$. The model, which is equivalent to the Jackiw-Teitelboim (JT) model first proposed in \cite{Jackiw:1984je, Teitelboim:1983ux}, can be viewed as arising from a dimensional reduction of an action where the dilaton plays the role of volume of the transverse space and goes to a constant in the IR. In the UV, the dilaton solution grows and regulates the backreaction allowing for finite-energy states. By computing the boundary correlation functions of an operator dual to a matter field, it was shown that the $AdS_2$ isometries are not respected in the IR. In particular, the classical four point function deviates away from conformality below a certain `breaking scale', $E_{br}$, which scales as $\sim G/V$, where $G$ is the higher dimensional Newton's constant and $V$ is the volume of the compact space $X$. Moreover, it becomes singular in the limit where the dilaton becomes constant reflecting the effect of backreaction in pure $AdS_2$.

The existence of this breaking scale resolves the puzzle with the density of states as it implies that the AdS$_2$ scaling symmetry is broken for energies below $E_{br}$. This means that the scaling argument used to derive \eqref{density} does not apply for low energy states, precluding the $1/E$ term.

Another peculiar feature of extremal black holes is the behavior of their thermodynamics upon heating them up slightly. Working in the canonical ensemble, one finds that their energy above extremality scales as $\alpha T^2$, for some scale $\alpha$ proportional to $G_N^{-1}$. This result suggests the presence of a critical scale $M_{gap} = 1/\alpha$ below which the total energy of the black hole is smaller than the temperature of the system \cite{Preskill:1991tb, Maldacena:1996ds, Maldacena:1997ih}. Below $M_{gap}$, or its `mass gap', the black hole does not have sufficient energy to emit a thermal quantum signaling the breakdown of the usual process of Hawking radiation. 

In this paper, we present evidence that the breaking scale of an extremal black hole coincides with its mass gap, $E_{br} \sim M_{gap}$. We check this for a wide range of examples including extremal BTZ and spherical/planar AdS Reissner-Nordstrom in any dimension. Furthermore, we show how the model of  \cite{Almheiri:2014cka} universally describes the near horizon geometry of extremal black holes and use it to prove that $E_{br} \sim M_{gap}$ holds generally. We emphasize that this agreement is noteworthy given the presence of many scales in the problem and that they are calculated from very different considerations. 

This paper is organized as follows. In section \ref{thermogap}, we review the notion of the thermodynamic mass gap. In section \ref{APreview}, we review the JT model studied in \cite{Almheiri:2014cka} and reproduce their four point function using another method that involves the computation of bulk Feynman diagrams. In section \ref{comparison}, we explicitly compute the breaking scale and mass gap for a large class of (near-)extremal black holes and show that they agree. In section \ref{universality}, we argue that the JT model provides a universal description of the IR physics of (near-)extremal black holes, and, using that fact, prove that the breaking scale and mass gap will always coincide. In section \ref{discussion}, we summarize our results and discuss their implications.

While this work was in preparation, related ideas were discussed from different perspectives in \cite{Jensen:2016pah, Maldacena:2016upp, Engelsoy:2016xyb}.

\section{The Thermodynamic Mass Gap of Near-extremal Black Holes} \label{thermogap}

Near-extremal black holes have this peculiar property that their semi-classical description seems to breakdown even while being macroscopic in size. As described in \cite{Preskill:1991tb}, since the total mass above extremality, $\Delta M = M - M_{ext}$ (at fixed charge), scales with temperature as
\begin{align}
\Delta M \simeq M_{gap}^{-1}T^2,
\end{align}
for some scale $M_{gap}$, it decreases faster than its temperature as $T \rightarrow 0$. Therefore, below the scale $M_{gap}$, the black hole will not have enough energy to emit the next Hawking quantum with typical energy $T$. Therefore, the semiclassical analysis of Hawking must breakdown. Following conventions in the literature we will call this scale the `thermodynamic mass gap' or simply its mass gap.

This conclusion is clearly dependent on how the energy scales with temperature. We show in section \ref{universality} that all near-extremal black holes in the canonical ensemble behave this way, but this conclusion can be arrived from more general considerations\footnote{It is important that the energy is defined for fixed charges and  not fixed chemical potentials. In the latter case, $\Delta M$ generically has a linear term in $T$ at low temperatures. We thank Blaise Gout\'eraux for pointing this out. For further discussions of this point, see appendix \ref{appendixC}.}. To see this, consider the specific heat of the black hole at fixed charge, $C_Q$. From the first law of thermodyamics we have
\begin{equation} \label{3rd}
S(T)-S(0) = \int^T_0 C_Q {dT \over T}.
\end{equation}
$S(T)$ is the entropy at temperature $T$. Since for a finite system the LHS has to be finite, $C_Q$ must vanish as $T \rightarrow 0$, and if it goes like $T^{\alpha - 1}$ at low temperatures, $\alpha$ has to be strictly greater than 1. If we further assume that $C_Q$ is an analytic function of $T$ around $T=0$, then the smallest $\alpha$ can be is 2. Therefore, in this case, the leading term of $\Delta M$ at low temperatures is generically expected to be quadratic in $T$\footnote{One may ask whether there are systems where $\alpha$ is not necessarily an integer which usually appears for systems with Lifshitz scaling or hyperscaling violation. In fact, as we discuss in section \ref{universality}, when the dominant saddle at zero temperature is a macroscopic extremal black hole, the near-extremal near-horizon geometry is still given by $AdS_2$ times some transverse manifold, $S(T) \propto T$, and in the canonical ensemble $E(T) \propto T^2$ as well.  For further discussions about this, see appendix \ref{appendixC}.}.

It is tempting to interpret this literally as the mass gap in the spectrum of  black hole masses in a fixed charge sector (up to a numerical factor of order one). This interpretation, though, does not quite follow from the above argument alone. \eqref{3rd} says that, as the temperature increases from zero to $M_{gap}$, the black hole entropy increases by order one bit.  But, since the number of states is the exponential of the entropy, this of course does not imply that there is necessarily a gap of order $M_{gap}$ in the spectrum\footnote{\cite{Page:2000dk} discusses various alternatives for the low-lying spectrum of near-extremal black holes, given the existence of the thermodynamic mass gap.}.

For some examples where a microscopic description is available \cite{Maldacena:1996ds}, it is possible to see explicitly that the thermodynamic mass gap is truly a mass gap of the spectrum. This mass gap, which is much smaller than the inverse of the effective size of the system, arises due to the twisted sectors of the microscopic theory describing the black hole, at least from the weak coupling point of view.  \cite{Maldacena:1997ih} also argued this using a very different method not depending on microscopic details of the theory. However, they implicitly assumed that the first excited state of an extremal Reissner-Nordstrom black hole is an extremal Kerr-Newman black hole. It is not clear whether this has to be the case as there could be lower energy states with zero angular momentum.

There is a caveat in the above argument for large black holes in asymptotically $AdS$ spacetimes. In this case, even below the mass gap, an incoming flux of energy on horizon can be in an equilibrium with the outgoing flux of energy, or Hawking radiation, and the outgoing flux of energy need not be constrained by $\Delta M$. If we seriously take the hypothesis that there is one single degree of freedom per Planck area on the black hole horizon \cite{'tHooft:1993gx, Susskind:1994vu}, then the horizon area should be quantized in units of the Planck area, and we could argue from \eqref{3rd} that there is a gap in the spectrum separated from the ground state by $M_{gap}$. As noted above, however, there is in general no reason why the entropy has to be quantized like this. Besides, it is not clear how one measures the area of a black hole to the precision of a single Planck area given that the fluctuations are usually of the same order. A probably stronger argument can be made if we take the AdS black hole slightly out of equilibrium for some amount of time so that there is an imbalance between the incoming and outgoing flux of energy. By allowing the black hole to evaporate, we will run into the same problem as above.

\section{The JT Model Revisited} \label{APreview}

\subsection{Review of the JT Model}

The JT model studied in \cite{Almheiri:2014cka} is a particular example of dilaton gravity with action
\be 
S = {1\over 16\pi G} \int d^2x \sqrt{-g} (\Phi^2R - U(\Phi)) + S_{matter}, \label{APaction}
\ee
where $U(\Phi) = -C(\Phi^2-\Phi_0^2)$ for some constants $C, \Phi_0^2>0$. For simplicity, $S_{matter}$ is taken to be the free massless scalar action in two dimensions. The zero-temperature solution of this action captures many of the important features of extremal black holes in higher dimensions. One can think of it as a dimensional reduction of an extremal black hole  down to two dimensions where the dilaton now represents the volume of the transverse directions. The zero-temperature solution is given by
\begin{align}
ds^2 &= {2 \over Cz^2}(-dt^2 + dz^2), \label{zero-temp metric} \\
\Phi^2 &= \Phi_0^2 + {a \over z}, \label{zero-temp dilaton}
\end{align}
where $a$ is a non-negative length scale that parameterizes a family of solutions.\footnote{Note that the parameter `a' in this paper is half the parameter `a' in \cite{Almheiri:2014cka}.} The role of the parameter $a$ is to regulate the backreaction in order to allow for asymptotic $AdS_2$ solutions with non-zero energy. As $a \rightarrow 0$, the dilaton becomes constant everywhere and no finite energy state can exist with the $AdS_2$ asymptotics.

The main result of \cite{Almheiri:2014cka} is to compute the effect of backreaction on boundary correlation functions of the operator dual to the scalar field. One might naively expect that the correlation functions can be computed by working in the probe limit and have their form be constrained by the conformal symmetry of $AdS_2$. In particular, since the scalar field is massless, it is dual to a dimension one operator necessitating that the correlation function $\langle \mO^n \rangle \sim 1/t^n$. However, when taking into account the backreaciton of the matter on the geometry, the classical four point function was found to scale as $\langle \mO^4 \rangle \sim 1/t^4 + G/at^3$, where the first piece is the disconnected contribution. The second term arises due to backreaction, as evident from the presence of $G$. Notice that in the $a \rightarrow 0$ limit, the four point function diverges, reflecting the problem of having non-trivial dynamics in $AdS_2$ with constant dilaton. At non-zero $a$ we see that there is a scale $E \sim G/a$ below which the correlators do not display the expected scaling symmetry.

The method used in \cite{Almheiri:2014cka} to compute the boundary correlation functions was to evaluate the on-shell action to obtain the boundary generating functional. Due to the backreaction of the boundary scalar sources on the dilaton, the near-boundary asymptotics of the dilaton becomes dependent on the sources, and one has to remove this dependence by redefining the boundary time to maintain the sourceless asypmtotics. It is in this new time coordinate that the generating functional becomes non-Gaussian and produces the four point function that breaks the conformal symmetry.

This method gives the impression that the loss of the conformal symmetry is a near-boundary or UV effect in contrast to the actual result as seen in the behavior of the four point function. We seek another method which makes the IR nature of the backreaction more manifest. Another disadvantage of this method is its technical difficulty when applied to more general realistic systems; The interpolation between the UV and IR geometries is usually not as simple as in the JT model, \eqref{zero-temp metric}, \eqref{zero-temp dilaton}, .

\subsection{A Bulk Linearized Quantum Field Theory}
\label{BQFT}
Here we introduce another method which overcomes the issues with the procedure of the previous subsection and is easily generalizable for other systems. The basic idea is to linearize the bulk fields and compute the correlation functions by evaluating bulk Feynman diagrams. As a check, we should be able to reproduce the four point function of \cite{Almheiri:2014cka} from a tree level diagram involving a graviton exchange.

We first need to choose the background on which to linearize. We use one diffeomorphism to gauge-fix the dilaton to have the form \eqref{zero-temp dilaton}, and the other to fix the metric to be diagonal. Our ansatz for the metric  and dilaton is
\begin{align}
ds^2&=e^{h_0}(-e^{h+g}dt^2+e^{h-g}dz^2),\\
e^{h_0}&={2 \over Cz^2},\\
\Phi^2&= \Phi_0^2 + {a\over z}. 
\end{align} 
where $h$ and $g$ will be treated as two linearized graviton perturbations. Plugging this into the action \eqref{APaction} gives
\begin{align}
S &= {1 \over 16 \pi G} \int dt dz \{\Phi^2 \partial_t^2\left( e^{-g} \right) - \Phi^2 \partial_z^2\left( e^{g} \right)   + \Phi^2 \partial_t \left( e^{-g} \partial_t (h_0 + h) \right) - \Phi^2 \partial_z \left( e^{g} \partial_z (h_0 + h) \right) \nonumber \\ & \ \ \ \ \ \ \ \ \ \ \ + C(\Phi^2 -\Phi_0^2)e^{h_0 + h}\} + \int dt dz \left\{e^{-g} {(\partial_t f)^2 \over 2} - e^{g} {(\partial_z f)^2 \over 2}\right\},
\end{align}
where $f$ is the linearized scalar perturbation about the vacuum. 

Next, we expand the action in $h$ and $g$. The zeroth order terms are simply irrelevant constants, while the linear order terms vanish by virtue of the background satisfying the equations of motion. At the quadratic level, we find that the graviton kinetic term is not diagonal in the fields $h$ and $g$. This can easily be amended with the field redefinition $h \rightarrow h - \partial_z \left( a \over z^2 g \right)/{2 a \over z^3}$. Focusing only on the terms which contribute to the classical four point function, the relevant part of the action is\footnote{Working in a fixed gauge will introduce Fadeev-Popov ghosts which, however, will not affect the classical connected four point function since they do not directly couple to the matter field.}
\begin{align}
S =  \int dt dz \left( -{a \over 64 \pi G z} \partial_z g \partial_z g - {1 \over 2}\eta^{i j}\partial_i f \partial_j f - {1\over 2}  \delta^{i j} \partial_i f \partial_j f   \ g   + ...\right) \label{APgraviton},
\end{align}
where we redefined the scalar field to absorb the $1/16 \pi G$. Suppressed here are higher point interactions between the graviton and the scalar field as well as self interactions of the graviton.

From the first term of this action we obtain the graviton propagator,
\be \label{gravitonpro}
G_g(z,t;z',t') = -{i 16 \pi G \over a}[z^2\Theta(z'-z)+z'^2\Theta(z-z')]\delta(t-t').
\ee    
Notice that this propagator is instantaneous in time, due to the absence of time derivatives in the kinetic term. This means that the field is not a propagating degree of freedom. Nevertheless, when coupled to matter fields, this propagator mediates their backreaction. 

To compute the four point function we also need the bulk-to-boundary propagator for the scalar field and the interaction vertex. Taking a limit of the bulk-to-bulk propagator, we obtain
\be
K(z,t;t') = {1\over \pi}{z \over z^2 - (t-t')^2},
\ee
and the interaction vertex is read off from the action to be
\be
V_{ffg}= -i(\partial_t^1\partial_t^2+\partial_z^1\partial_z^2),
\ee
where the subscripts 1 and 2 refer to the two incoming scalar fields. Putting all these ingredients together we find that the four point function, after adding all three $s,t,$ and $u$ channels, is given by
\begin{align}
&A_4(t_1,t_2,t_3,t_4) =\nonumber \\
&- \int d^2xd^2x' \{\partial_tK(z,t;t_1)\partial_tK(z,t;t_2)+\partial_zK(z,t;t_1)\partial_zK(z,t;t_2)\} 
 \nonumber \\
& \ \ \ \ \ \ \ \ G_g(z,t;z',t')\{\partial_{t'}K(z',t';t_3)\partial_{t'}K(z',t';t_4)+\partial_{z'}K(z',t';t_3)\partial_{z'}K(z',t';t_4)\} \nonumber \\
& + (t_2 \leftrightarrow t_3) +(t_2 \leftrightarrow t_4) \label {4pt_integral}.
\end{align}
We were able to compute the above integral explicitly for the special case of time arrangements $\{ t_1, t_2, t_3, t_4\} = \{ \Delta + \delta, \Delta, \delta, 0 \}$ with $\Delta > \delta > 0$, and obviously with any overall time shift, and we found exact\footnote{Up to a factor of 2. In fact, the Schwarzian term in the gravitational on-shell action missed in \cite{Almheiri:2014cka} exactly accounts for this factor of 2 discrepancy. We thank Kristan Jensen, Juan Maldacena, and Zhenbin Yang for pointing this out to us.} agreement with the result of \cite{Almheiri:2014cka}. We also evaluated this integral numerically and found the same agreement for arbitrary times. This gives strong credence to the bulk linearized field theory approach.

From this perspective, the conformal symmetry breaking is manifestly an IR effect; the graviton propagator scales as $z^2$ and so the diagram receives most of its contribution in the IR. Moreover, since the four-point function is proportional to $G/a$ and the scalar operator has mass dimension one, one anticipates simply from dimensional analysis that it should scale as $\sim G/a t^3$. Indeed, one can explicitly check that when $t_i \rightarrow \lambda t_i$, the integral \eqref{4pt_integral} scales as $\sim G/a\lambda^3$.

\subsection{Thermodynamics and mass gap in the JT model}
In this subsection, we review the thermodynamics of the JT model and compute the mass gap of the theory. We will show that it occurs at the same scale as $E_{br}$.  Consider the finite temperature solution
\begin{align}
ds^2 &= {4(\mu/a) \over C\sinh^2[\sqrt{2\mu/a}z]}(-dt^2+dz^2), \label{finite-temp metric}\\
\Phi^2 &= \Phi_0^2 + \sqrt{2\mu a}\coth[\sqrt{2\mu/a}z], \label{finite-temp dilaton}
\end{align} 
where $\mu$ is a mass scale that determines the temperature and mass of the solution. As $\mu \rightarrow 0$, 
\eqref{finite-temp metric}, \eqref{finite-temp dilaton} reduces to \eqref{zero-temp metric}, \eqref{zero-temp dilaton}. The temperature $T$ and mass $E$ of the above solution are
\begin{align}
T &= {1 \over \pi}\sqrt{\mu \over 2a},\\
E &= {\mu \over 8\pi G}.
\end{align}
In particular, $E$ depends on $T$ as
\be \label{APthermo}
E = {\pi a \over 4G}T^2, 
\ee
which gives
\be
M_{gap} = {4G \over \pi a}.
\ee
Therefore, $M_{gap} \sim E_{br} \sim G/a$ in the JT model, consistent with our general claim. In fact, as we will show in section 5, the JT model is a universal description of near-horizon physics of near-extremal black holes. The fact that $M_{gap} \sim E_{br}$ in the JT model then guarantees that the same holds true for other more realistic theories. 

\section{Comparison of the Mass Gap and Conformal Symmetry Breaking Scale} \label{comparison}

As mentioned in the end of the last section, the fact that $M_{gap} \sim E_{br}$ in the JT model, together with the universality of the JT model, in principle proves the equivalence in general. Nonetheless, it is both an instructive exercise and a useful consistency check to directly compute and compare the scales of the mass gap and conformal symmetry breaking in a large class of near-extremal black holes.

As outlined in the previous sections, the mass gap of a near-extremal black hole can be easily read off from the low temperature expansion of the energy above extremality. The conformal symmetry breaking scale, we argue, can be read off from the scale appearing in the graviton propagator in the IR, as demonstrated in section \ref{BQFT}. To apply the same reasoning in the general case, we assume when computing the IR limit of the tree level diagram involving the exchange of a graviton that the matter bulk-to-boundary propagator can be approximated by its $AdS_2$ conformal form. We find detailed agreement between these scales.

\subsection{Extremal BTZ black holes}

We begin with the case of an extremal BTZ black hole in $2+1$ dimensions. The metric of  a BTZ black hole is given by \cite{Banados:1992wn}
\begin{equation}
ds^2 = -{(r^2-r_+^2)(r^2-r_-^2) \over l^2r^2}dt^2 + {l^2r^2 \over (r^2-r_+^2)(r^2-r_-^2)}dr^2 + r^2(d\phi+{r_+r_- \over r^2}dt)^2.
\end{equation}
Its mass, angular momentum, and temperature are
\be
M={r_+^2+r_-^2 \over 8Gl^2}, \ \ \ J={r_+r_- \over 4Gl^2}, \ \ \ T={r_+^2-r_-^2 \over 2\pi l^2 r_+}.
\ee
To compute the mass gap of the near-extremal BTZ black hole with fixed angular momentum $J = r_0^2/ 4Gl^2$, we vary $(r_+ - r_-)$ while maintaining $r_0^2 = r_+r_-$.
We define the energy above extremality to be
\begin{equation}
\Delta E \equiv M-J = {(r_+ - r_-)^2 \over 8Gl^2}.
\end{equation}
The temperature near extremality becomes
\begin{equation}
T \simeq {(r_+ - r_-)2r_0 \over 2\pi l^2 r_0} = {r_+ - r_- \over \pi l^2},
\end{equation}
and so
\begin{equation}
\Delta E \simeq {\pi^2 l^2 T^2 \over 8G}.
\end{equation}
From this relation we can read off the mass gap, ignoring numerical factors, as
\begin{equation} \label {btzgap}
M_{gap} \sim {G/l^2}.
\end{equation}

Next we turn to the computation of the conformal symmetry breaking scale of the extremal BTZ black hole. To this end, we consider the action of three dimensional Einstein gravity plus a massless scalar field,
\begin{equation}
S=\frac{1}{16\pi G}\int d^3x\sqrt{-g_3}\left(R_3+\frac{2}{l^2}\right)  -\frac{1}{2}\int d^3x\sqrt{-g_{(3)}}g_{(3)}^{\mu\nu}\partial_{\mu}f \partial_{\nu}f.
\end{equation}
We wish to focus on the s-wave sector of this theory and therefore cast it in terms of its two dimensional truncation using the following ansatz
\begin{equation} \label{btz-reduction}
ds_3^2 = g_{\mu\nu}dx^{\mu}dx^{\nu} + e^{-2\psi}l^2(d\phi+A_{\mu}dx^{\mu})^2,
\end{equation}
where $\phi$ has period $2\pi$. After this dimensional reduction, $g_{\mu\nu}, \psi$, and $A_{\mu}$ respectively become the metric, dilaton, and gauge fields of the two dimensional theory. The action for the gravitational sector becomes \cite{Strominger:1998yg, Castro:2008ms}
\begin{equation} \label{btz-action}
S_{grav}=\frac{l}{8G}\int d^2x\sqrt{-g}e^{-\psi}\left(R+\frac{2}{l^2}-\frac{l^2}{4}e^{-2\psi}F^2\right).
\end{equation}
We will study perturbations around the extremal BTZ black hole, which has a near-horizon $AdS_2$ region, with the metric
\begin{equation}
ds_3^2=-\frac{(r^2-r_0^2)^2}{l^2r^2}dt^2+\frac{l^2r^2}{(r^2-r_0^2)^2}dr^2+r^2(d\phi+\frac{r_0^2}{lr^2}dt)^2.
\end{equation}
The background values of the two dimensional fields are then
\begin{align}
ds^2&= - e^{g_0}dt^2 + e^{-g_0}dr^2 \label{btz-bgd-metric}\\
e^{-2\psi}&=\frac{r^2}{l^2}\label{btz-bgd-dilaton}\\
\bar{A}_{\mu}dx^{\mu}&=\frac{r_0^2}{lr^2}dt, \label{btz-bgd-gauge}
\end{align} 
where $e^{g_0}=(r^2-r_0^2)^2/l^2r^2$. As $r \rightarrow r_0$, the two dimensional metric approaches $AdS_2$, 
\be \label{btzads2}
ds^2 \rightarrow -{4 (r-r_0)^2 \over l^2} dt^2 + {l^2 \over 4(r-r_0)^2} dr^2,
\ee
where the $AdS_2$ radius is $l/2$. We want to linearize about this background and consider the bulk linearized quantum field theory which couples the metric, gauge, and scalar perturbations. Consider first the action for the scalar perturbations
\begin{align}
S_{f} = -{1 \over 2} \int d^3 x \sqrt{- g_{(3)}} g_{(3)}^{\mu \nu} \partial_\mu f \partial_\nu f.
\end{align}
When dimensionally reducing, we assume that $f$ has no dependence on the transverse dimension $\phi$, so we can set those derivatives to zero. Another simplification is offered by working in the gauge $A_r = 0$ which implies no coupling between the scalar and the gauge field. Consequently,
\begin{align}
g_{(3)}^{\mu\nu}\partial_{\mu}f \partial_{\nu}f = g^{ab}\partial_{a} f \partial_{b} f.
\end{align}
Since the three dimensional determinant reduces to $\det g_{(3)}= e^{-2\psi}l^2\det g$, the reduced scalar action becomes
\begin{align}
S_{f} = -{V_{\phi} \over 2} \int d^2 x \sqrt{- g} l e^{-\psi} g^{a b} \partial_a f \partial_b f, \label{scalar-btz-action}
\end{align}
where $V_{\phi}=  2 \pi$ is the coordinate volume of the transverse direction $\phi$.

Before moving on to consider the other fields, we first discuss the expected IR behavior of boundary correlations functions ignoring the effects of backreaction. Let's focus on the two point function of the operator dual to the scalar field. Notice first that the scalar action above and that of a free massless scalar in $AdS_2$ differ by the presence of the dilaton term $e^{-\psi}$ and the background metric. However, in the near-horizon region,  the dilaton goes to a constant and the metric approaches $AdS_2$. Therefore, one should expect that correlation functions which probe the IR geometry, namely those with large boundary time separations, should transform covariantly under the IR $AdS_2$ scaling symmetry, and thus should scale as $\sim 1/t^{2}$ .

A more direct way of seeing this emergent IR symmetry is to directly compute the two point function in the extremal BTZ background and take the long time limit. Without the periodic identification of $\phi$, the extremal BTZ metric is diffeomorphic to the vacuum $AdS_3$. Therefore, the two point function in this case can be obtained from the two point function in the vacuum state by a conformal transformation \cite{KeskiVakkuri:1998nw},
\be \label{noncompact}
\langle O(t,\phi)O(0,0) \rangle = {\exp(-{r_0\Delta\over l}(\phi-t/l)) \over \left[(1-\exp(-{2r_0\over l}(\phi-t/l)))(\phi+t/l)\right]^\Delta}.  
\ee
Note that \eqref{noncompact} decays exponentially in time, even though the IR geometry is $AdS_2$. This is because the transverse direction is infinite so that no compact perturbation from the boundary truly becomes an s-wave, even in the long time limit.

The two point function on the extremal BTZ background after periodically identifying $\phi$ is obtained by the method of images in the bulk, 
\begin{align} \label{compact}
&\langle O(t,\phi)O(0,0)\rangle = \sum^{\infty}_{n=-\infty} {\exp(-{r_0\Delta\over l}(\phi-t/l+2\pi n)) \over \left[(1-\exp(-{2r_0\over l}(\phi-t/l+2\pi n)))(\phi+t/l+2\pi n)\right]^\Delta}.
\end{align}
We focus on the case of a massless scalar field in three dimensions with $\Delta = 2$. In this case, all the terms in the sum are manifestly positive and each is exponentially suppressed in $\phi - t/l + 2 \pi n$ times a power law which goes as $1/(\phi + t/l + 2 \pi n)^{2}$. We care about the largest  contribution at late times. This sum will be dominated by the term which is least exponentially suppressed and thus with the minimal $\phi - t/l + 2 \pi n$. In fact, for generic $t>0$ we can always find an $n$ such that $\phi - t/l + 2 \pi n \sim O(1)$, where $2 \pi n \sim t/l$. More precisely, we can always find $n$ such that $|-t/l + 2\pi n| < 2\pi$. Since such a term exists for any large $t$, we conclude that the two point function decays approximately as a power law and goes as $1/t^{2}$. This is precisely the behavior one expects for a massless scalar field in $AdS_2$. The same story holds true for general $\Delta$. That is, in the long time limit, the two point function of a general massive scalar field in the extremal BTZ scales like that of the scalar field with the same mass in $AdS_2$.\footnote{The above argument suggests that at least the bulk tree-level two point function respects the IR conformal symmetry as naively expected. But, since the IR conformal symmetry is actually explicitly broken as we argue below, it would be surprising if the exact two point function respects that symmetry. Indeed, the method of images in general does not work in the boundary theory, and  \eqref{compact} is therefore not expected to be the exact two point function.}

Now we will demonstrate how this expectation fails once backreaction is taken into account. First, we introduce fluctuations of the metric and gauge field about the background \eqref{btz-bgd-metric}, \eqref{btz-bgd-gauge} as
\begin{align} 
ds^2&=-e^{(g_0+g)+h}dt^2 + e^{-(g_0+g)+h}dr^2,\\
A_t &= \bar{A}_t + a.
\end{align}
Just as in the JT model, we work in a gauge where the dilaton is fixed to \eqref{btz-bgd-dilaton}. We also work in a guage where $A_r = 0$. With this ansatz, the quadratic part of the action \eqref{btz-action} becomes
\begin{align}
S_{quad}={l \over 8 G}\int dt dr \left[\left(-{2(r^4-r_0^4)\over l^3r^3}g -{(r^2-r_0^2)^2 \over l^3r^2}\partial_r g +{2r_0^2\over l^2}\partial_r a\right)h+{r^4+r_0^4 \over l^3 r^3}h^2 + {r^3\over 2l} (\partial_r a)^2\right].
\end{align}
Notice that this action is different from what we found for the JT model in that it couples the graviton to the gauge field. Since the graviton $g$ is what couples to the scalar field, we wish to find its propagator in the IR. To diagonalize the quadratic terms involving $g$, $h$, and $a$, we integrate out the latter two fields simply by plugging in their respective equations of motion back into the action. Their equations of motion are
\begin{align}
-{2r_0^2 \over l^2}\partial_r h -\partial_r\left[{r^3 \over l}\partial_r a\right]=0, \\
-{8r\over l^3}g - {4r^2\over l^3}\partial_r g +{8r \over l^3}h=0.
\end{align}
Solving these equations and plugging back in the solutions we find a quadratic action purely of $g$ to be
\begin{align}
S_{quad} &= -{1\over 8 G l^2}\int dt dr r^3 (\partial_r g)^2    \\
&= - \int dt dz {l^2 \over 128 G z} (\partial_z g)^2,
\end{align}
where in the last line we transformed to Poincare coordinates where $z = l^2/4(r - r_0)$. This has precisely the same form as \eqref{APgraviton} and so we can directly read off the breaking scale to be
\begin{align}
E_{br} \sim {G \over l^2},
\end{align}
which agrees with $M_{gap}$ in \eqref{btzgap}. We stress that this agreement was not guaranteed simply from dimensional analysis as there is another scale in the problem $r_0$ which does not appear.

\subsection{Spherical charged black holes in AdS}
Next, we consider the case of spherical charged black holes in AdS. We will consider electrically charged black holes in arbitrary dimensions, and the dyonic ones in $3+1$ dimensions. Consider first the Einstein-Maxwell theory in $AdS_{n+1}$ for $n \ge 3$ whose action is \cite{Chamblin:1999tk}
\begin{align}
S = {1 \over 16 \pi G} \int d^{n+1} x \sqrt{-g}\left[R + {n(n-1)\over l^2}-F^2\right].
\end{align}   
The charged black hole solution is given by
\begin{align}
ds^2 &= -V(r)dt^2+{dr^2 \over V(r)} + r^2 d\Omega_{n-1}^2,\\
\bar{A}_{\mu}dx^{\mu} &= -\sqrt{n-1 \over 2(n-2)}\left({q\over r^{n-2}}-{q\over r_+^{n-2}}\right)dt,
\end{align}
where
\begin{align}
V(r) = 1 - {m\over r^{n-2}}+{q^2 \over r^{2n-4}} + {r^2\over l^2},
\end{align}
and $r_+$ is its largest root.
Its asymptotic mass $M$, charge $Q$, and temperature $T$ are given by
\begin{align}
&M={(n-1)w_{n-1}\over 16\pi G} m,\\
&Q={\sqrt{2(n-1)(n-2)}w_{n-1} \over 8\pi G} q,\\
&T= {2 r_+^{2n-2}+m (n-2)l^2r_+^{2n-4}-2 (n-2)q^2l^2 \over 4\pi l^2 r_+^{2n-3}},
\end{align}
where $w_{n-1}$ is the volume of the $n-1$ unit sphere. We can solve $V(r_+) = 0$ for $m(r_+)$ and express the temperature as a function of $r_+$ and $q$
\begin{align}
T= {nr_+^{2n-2}+(n-2)l^2r_+^{2n-4}-(n-2)q^2l^2 \over 4\pi l^2 r_+^{2n-3}}.
\end{align}
When $T = 0$ the black hole becomes extremal and the location of its event horizon, $r_0$, is determined by the following equation
\begin{align}
\left({n\over n-2}\right)r_0^{2n-2}+l^2r_0^{2n-4} = q^2l^2. \label{t0eqn}
\end{align} 
The near-horizon geometry of this solution is given by
\begin{align}
ds^2 &= -{(r - r_0)^2 \over L^2}dt^2+{L^2 \over (r - r_0)^2}{dr^2} + r_0^2 d\Omega_{n-1}^2,
\end{align}
where $L^2 = \left( {n(n-1) \over l^2} + {(n-2)^2 \over r_0^2}  \right)^{-1}$ is the square of the $AdS_2$ radius.

We now compute the mass gap of the near-extremal black hole. We fix the charge of the black hole in terms of $r_0$ by solving \eqref{t0eqn} and then slightly increase its mass to give it a non-zero temperature at fixed charge. The new horizon radius increases to $r_+ \equiv r_0 + \delta$, and the mass increases from $M_{ext}$ to $M$. The latter can be expanded in terms of $\delta$ as
\begin{align}
&{16\pi G \over (n-1)w_{n-1}}(M-M_{ext})
= r_+^{n-2} + {q^2 \over r_+^{n-2}} + {r_+^n \over l^2} -\left(r_0^{n-2} + {q^2 \over r_0^{n-2}} + {r_0^n \over l^2} \right) \nonumber \\
&= \left[{1\over 2}(n-2)(n-3)r_0^{n-4}+{(n-1)nr_0^{n-2} \over 2l^2}+ {1\over 2}(n-1)(n-2)q^2r_0^{-n}\right]\delta^2 + O(\delta^3) \\
&= {r_0^{n-4} \over l^2} \left[ l^2 (n-2)^2 + n (n-1) r_0^2 \right] \delta^2 + O(\delta^2).
\end{align} 
Note that the linear term vanishes. Similarly,
\begin{align}
T &= {r_0^{-2n-2}(nr_0^{2n+2}-l^2(n-2)((3-2n)q^2r_0^4+r_0^{2n}))\over 4\pi l^2}\delta + O(\delta^2) \\
&= {l^2 (n-2)^2 + (n-1)n r_0^2 \over 2 \pi l^2 r_0^2} \delta + O(\delta^2)
\end{align}
Therefore, for small $\delta$,
\be
M-M_{ext} \simeq M_{gap}^{-1}T^2,
\ee
where
\begin{align}
M_{gap} &= {4 G [(n-2)^2l^2+n(n-1)r_0^2] \over \pi(n-1)w_{n-1} l^2 r_0^n} \\
 &= \left( {4  \over \pi (n-1) w_{n-1} }\right) {G \over L^2 r_0^{n-2}}    \label{Smassgap}
\end{align} 
where in the second line we re-expressed the result in terms of the $AdS_2$ radius.

The dyonic result follows from the previous analysis simply by plugging $n = 3$ and replacing $q^2 \rightarrow q_E^2 + q_B^2$, the sum of the squares of the electric and magnetic charges of the black hole respectively. Thus, the dyonic black hole mass gap is
\begin{align}
M_{gap}^{Dyonic} &= {2 G ( l^2+ 6 r_0^2 ) \over \pi w_{2} l^2 r_0^3}.  \\
&= \left( {2  \over \pi  w_{n-1} }\right) {G \over L^2 r_0}\label{mgdyonic}
\end{align}

Now we turn to the computation of the breaking scales of these black holes. We begin with the dyonic case. Consider the following ansatz for the background metric and its perturbations
\begin{align} \label{metric-ansatz}
ds^2 = -e^{h+g + g_0}dt^2 + e^{h-g - g_0}dr^2 + r^2d\Omega_2^2,
\end{align}
where $e^{g_0} = V(r)$ is the background metric. With this ansatz, the quadratic part of the gravitational action is
\begin{align}
\sqrt{-g}\left( R  + {6 \over l^2} \right) \rightarrow 2 \sin \theta \left[ {h^2 \over 2} \left( 1 + {3 r^2 \over l^2} \right) + r V(r) g \partial_r h \right]
\end{align}
Now we consider the Maxwell term. The field strength tensor expanded about a background is
\begin{align}
F^2 &= (\bar{F}+f)_{\mu\nu}(\bar{F}+f)^{\mu\nu} \nonumber \\
&= \bar{F}^2 + 2f_{ab}\bar{F}^{ab} + f_{ab}^2,
\end{align}
where $\bar{F}$ is the background, $f$ the fluctuation, and $a,b \in \{t,r\}$. These are the only allowed perturbations which respect the dimensional reduction; see appendix \ref{appendixA} for an argument. The background values for the gauge field and field strength are
\begin{align}
\bar{A}_\mu dx^{\mu} &= {q_E (r - r_+) \over r r_+} dt + q_M \cos\theta d\phi, \\
\bar{F}^2 &=-{2q_E^2 \over r^4}e^{-2h} + {2q_M^2 \over r^4}.
\end{align}
This gives the following contribution to the quadratic action
\begin{align}
\sqrt{-g}(-F^2)&\rightarrow -r^2\sin\theta\left[{h^2 \over r^4}(-q_E^2+q_M^2)-{4q_E \over r^2}f_{tr}h+f_{ab}^2\right].
\end{align}
Thus, the full quadratic action is
\begin{align}
S_{quad} = {1 \over 16 \pi G}\int dt dr 2 \sin \theta \left[ {h^2 \over 2} \left( 1 + {3 r^2 \over l^2}  + {q_E^2 - q_M^2 \over r^2}\right) + r V(r) g \partial_r h +{4q_E}f_{tr}h- r^2 f_{ab}^2 \right].
\end{align}
Just as before, it is only the metric perturbation $g$ that couples to the scalar field, and  so we integrate out all other perturbations. Working in the gauge where $f_{tr} = - \partial_r a_t$, we find the following equations of motion for $f_{t r}$ and $h$
\begin{align}
&-4 q_E \partial_r h + 4 \partial_r \left( r^2 \partial_r a_t \right) = 0, \\
&-\partial_r \left( r V(r) g \right) + h \left( 1 + {3 r^2 \over l^2}  + {q_E^2 - q_M^2 \over r^2} \right) = 0.
\end{align}
Solving these equations and plugging the solutions back in, we find the following quadratic action for the metric perturbation $g$
\be
S_{quad}=-{1 \over 4G}\int d^2x \left[{\left( \partial_r[rV(r)g] \right)^2\over 1 + {3 r^2 \over l^2}  -  {q_E^2 + q_M^2 \over r^2}}\right].
\ee
Since we are interested in the IR behavior of the graviton propagator, we expand the action in $\Delta r/r_0$ where $\Delta r = r - r_0$.  We regard $\partial_r$ as a negative power of $\Delta r$ when comparing the relative size of each term. Using the form of the charge and mass as a function of $r_0$
\begin{align}
&m = r_0 + {q^2 \over r_0} + {r_0^3 \over l^2},\\
&(q_E^2 + q_M^2) l^2 = 3r_0^4 + l^2r_0^2,
\end{align} 
we find 
\begin{align}
S_{quad}=-{1\over 4G}\int d^2 x {l^2 +6r_0^2 \over 2l^2 r_0}(r - r_0)^3(\partial_r g)^2.
\end{align}
Changing coordinates to Poincare $AdS_2$ via $1/z =  ({6 \over l^2} + {1 \over r_0^2})(r - r_0)$ the action becomes
\begin{align}
S_{quad}=-{1\over 8G}\int dtdz {r_0 \over ({6 \over l^2} + {1 \over r_0^2}) z}(\partial_z g)^2,
\end{align}
from which we can read off the breaking scale to be
\begin{align}
E_{br} \sim {G  ({6 \over l^2} + {1 \over r_0^2})  \over r_0} \sim {G \over L^2 r_0},
\end{align}
where $L$ is the $AdS_2$ radius. This is precisely the same scaling we found in \eqref{mgdyonic}.

Next, we consider the case of electrically charged black holes in $AdS_{n+1}$ that arise from the action
\begin{align}
S = {1\over 16\pi G}\int d^{n+1}x \sqrt{-g} \left[R+{n(n-1) \over l^2} -F^2\right].
\end{align}
We consider again the metric ansatz
\begin{align}
ds^2 = -e^{h+g+g_0}dt^2 +e^{h-g-g_0}dr^2 + r^2d\Omega_{n-1}^2.
\end{align}
We will write the metric determinant as $\sqrt{-g} = e^hr^{n-1}\sqrt{g_{n-1}}$, where $g_{n-1}$ is the determinant of the $n-1$ sphere metric. Following similar steps as above we find
\begin{align}
\sqrt{-g}R &\rightarrow \sqrt{g_{n-1}}\left[{(n-1)(n-2)r^{n-3} \over 2}h^2 + (n-1)r^{n-2}V(r)g\partial_rh\right],\\
\sqrt{-g}{n(n-1) \over l^2}  &\rightarrow \sqrt{g_{n-1}}{n(n-1)\over 2l^2}r^{n-1}h^2,\\
\sqrt{-g}(-F^2) &\rightarrow \sqrt{g_{n-1}}\left[{(n-1)(n-2)q^2 \over 2r^{n-1}}h^2  + 4 g \sqrt{(n-1)(n-2) \over 2} f_{t r} h - r^{n-1} f^2_{a b}   \right],
\end{align}
 After integrating out the gauge field and metric perturbation $h$ we end up with the action
\begin{align}
S_{quad} = -{w_{n-1}(n-1) \over 32 \pi G} \int d^2x \left[ {\left( \partial_r \left[  r^{n-2} V(r) g  \right] \right)^2\over   W(r) } \right]
\end{align}
where $W(r) = {(n-1)(n-2) \over 2}r^{n-3} + {n(n-1) \over 2l^2}r^{n-1} - {(n-1)(n-2) \over 2r^{n-1}q^2}$. After performing a near-horizon expansion and transforming to Poincare coordinates, we find
\be
S_{quad}=-{w_{n-1}(n-1)\over 64\pi G}\int dtdz {r_0^{n} l^2  \over \left( (n - 2)^2 l^2 + n (n-1) r_0^2 \right) z} (\partial_z g)^2.
\ee
Again, we find agreement with the mass gap in \eqref{Smassgap} and the breaking scale,
\begin{align}
E_{br} \sim {G\left[ (n - 2)^2 l^2 + n (n-1) r_0^2 \right]  \over r_0^{n} l^2 } \sim {G \over L^2 r_0^{n-2}}.\label{EbrSph}
\end{align}

\subsection{Planar charged black holes in AdS}
Next, we consider the case of planar black holes in AdS. One way of obtaining these black holes is by starting with the spherical black hole solutions of the previous subsection and taking a scaling limit whereby the radius of the black hole is taken to infinity. The planar black hole obtained in this way has an infinite transverse volume. This infinite volume renders the effective Newton's constant zero thus trivializing the effect of backreaction; both the mass gap and breaking scale vanish in this case. Instead, we consider the situation where the transverse directions are compactified on a torus. We will be brief in this section as the steps are very similar to those in the spherical case. We begin with the same action keeping in mind the topology of the transverse space,
\begin{align}
S = {1 \over 16 \pi G} \int d^{n+1} x \sqrt{-g}\left[R + {n(n-1)\over l^2}-F^2\right].
\end{align}   
The background solution is
\begin{align}
ds^2 &= -U(r)dt^2 + {dr^2 \over U(r)} +r^2dx_i^2,\\
\bar{A}_{\mu}dx^{\mu} &= -\sqrt{n-1 \over 2(n-2)}\left({q\over r^{n-2}}-{q\over r_+^{n-2}}\right)dt,
\end{align}   
where
\begin{align}
U(r) = {r^2 \over l^2} -{m \over r^{n-2}} + {q^2 \over r^{2n-4}}. \label{uplanar}
\end{align}
Its mass $M$, charge $Q$, and temperature $T$ are \cite{Faulkner:2009wj}
\begin{align}
M&={(n-1) \over 16\pi G} mV,\\
Q&={\sqrt{2(n-1)(n-2)} \over 8\pi G}qV,\\
T&={2 r_+^{2n-2}+ (n-2)l^2 m r^{n-2}-(2n-4)q^2l^2 \over 4\pi l^2 r_+^{2n-3}},
\end{align}
where $V$ is the (dimensionless) coordinate volume of the transverse directions $x_i$. Using \eqref{uplanar} to solve for $m(r_+)$, we express the temperature as a function of $r_+$ and $q$ 
\begin{align}
T&={nr_+^{2n-2}-(n-2)q^2l^2 \over 4\pi l^2 r_+^{2n-3}},
\end{align}
Expanding the temperature and mass for a near-extremal black hole about extremality and following essentially the same steps as in the previous section, we find
\be \label{Pmassgap}
M_{gap} = {4nG \over l^2 Vr_0^{n-2}} = {4G \over (n-1)L^2 Vr_0^{n-2}},
\ee
where $L^2 = l^2/n(n-1)$ is the IR $AdS_2$ radius. As in the spherical case, we obtain the dyonic black hole result by replacing $q^2$  by $q^2_E+q^2_M$, with the gauge potential being modified to
\be
\bar{A}_{\mu}dx^{\mu} = -\left({q_E \over r} - {q_E \over r_+}\right)dt - (q_M y)dx,
\ee
but the mass gap is still given by \eqref{Pmassgap}.

Now we compute the breaking scale of these planar black holes. We use the same ansatz as \eqref{metric-ansatz} and working in the gauge $f_{t r} = - \partial_r a_t$. Integrating out all the fields except for the graviton $g$, which couples to the scalar, we find the action
\be
S_{quad}=-{V(n-1) \over 32 G} \int d^2x {\left( \partial_r \left[r^{n-2}U(r)g \right]\right)^2\over W(r)}
\ee
where we defined
\begin{align}
W(r) \equiv {n(n-1)r^{n-1} \over 2l^2} - {(n-1)(n-2)q^2 \over 2r^{n-1}}.
\end{align} 
In case of the dyonic black hole, $q^2$ above should be replaced with $q_E^2 + q_M^2$.  In the near-horizon region, the action becomes
\be
S_{quad} =-{V \over 64 \pi G} \int dtdz {l^2r_0^{n-2} \over n}{1\over z} (\partial_r g)^2,
\ee
in terms of the Poincare coordinate $z$ of the IR $AdS_2$. 
Thus, the breaking scale is
\begin{align}
E_{br} \sim {nG \over l^2Vr_0^{n-2}}  = {G \over (n-1)L^2 Vr_0^{n-2}}.
\end{align}
and agrees with the mass gap \eqref{Pmassgap}.

\section{Universality of the JT model} \label{universality}

There is a sense in which the JT model gives a universal description of the near-horizon $AdS_2$ region of extremal black holes in a large class of dilaton gravity theories.\footnote{We thank Douglas Stanford for suggesting this possibility.} 
Consider a dilaton gravity theory whose action is given by
\be \label{realistic}
S = {1\over 16\pi G_2} \int d^2x \sqrt{-g} (\Phi^2R +\lambda(\nabla \Phi)^2- U(\Phi) - f(\Phi)F^2),
\ee
for some $\lambda, U(\Phi)$ and $f(\Phi)$, where $F^2$ is a Maxwell term. This can be viewed as a dimensional reduction of a $(n + 1)$-dimensional theory where $\Phi^2$ is the coefficient of the transverse metric raised to the power of $(n-1)/2$. The total volume of this space, $X$, is $\Phi^2 V_X$ where $V_X$ is its coordinate volume. 
The action \eqref{realistic} describes the dimensional reduction of a large class of well known higher dimensional theories including Einstein-Maxwell theory, for which $\lambda = 4(n-2)/(n-1)$. By a Weyl transformation
\be \label{Weyl_trans}
g_{a b} \rightarrow g_{a b} \Phi^{-\lambda /2},
\ee
one can set $\lambda =0$ with $U(\Phi) \rightarrow \Phi^{-\lambda/2}U(\Phi)$ and $f(\Phi) \rightarrow \Phi^{\lambda/2}f(\Phi)$. Since we are interested in on-shell quantities, one can further eliminate the Maxwell term by solving the gauge field equations of motion and plugging back its solution, as explained in appendix \ref{appendixB}, assuming no charged matter. Therefore, without loss of generality, we will assume $\lambda = 0$ and $f(\Phi) = 0$. 

Next, we look for a general static solution of this action. Working in the gauge where the 2 dimensional metric is 
\begin{align}
ds^2 = - e^{2 w} dt^2 + e^{- 2 w} dr^2,
\end{align}
the equations of motion become
\begin{align}
&  2 w' (\Phi^2)'   +  (\Phi^2)''         + e^{-2 w} U(\Phi) = 0, \\
&4 (w')^2 + 2 w'' + e^{-2 w} \partial_{\Phi^2}U(\Phi) = 0, \\
&(\Phi^2)' = -{\eta \over 2}, \label{dilgeneq}
\end{align}
for some $\eta$ which parametrizes a family of solutions. Equation \eqref{dilgeneq} gives $\Phi^2 = \Phi^2_h - {\eta \over 2} r$, which when plugged back in gives the following differential equation for the metric
\begin{align}
\left( e^{2 w} \right)' = {2 \over \eta}U(\Phi).
\end{align}
For general $U(\Phi)$, we can look for solutions near the point where $\Phi^2 = \Phi^2_h$ by Taylor-expanding in $\eta r$. The differential equation becomes
\begin{align}
\left( e^{2 w} \right)' = {2 \over \eta} \left( U(\Phi_h) + \partial U(\Phi_h) \left({-\eta \over 2}\right) r + {1 \over 2}\partial^2 U(\Phi_h) \left({-\eta \over 2}\right)^2 r^2 + \cdots  \right), \label{fulle2w}
\end{align}
where $\partial U \equiv \partial_{\Phi^2} U$. Notice that truncating this expansion at first nontrivial order in $\eta r$ would give the equation of motion of the JT model; recall that $\partial^{n>1} U_{AP}(\Phi) = 0$. Therefore, this demonstrates that the JT model correctly captures the near-horizon physics of (near-)extremal black holes at tree level.

Integrating \eqref{fulle2w}, the static solution near $\Phi_h^2$ is given by
\begin{align}
&\Phi^2 = \Phi^2_h - {\eta \over 2} r, \\
&ds^2 = - f(r) dt^2 + {dr^2 \over f(r)},
\end{align}
where
\begin{align}
f(r) = e^{2 w_0} + {2 \over \eta} \left( U(\Phi_h) r + {1 \over 2}\partial U(\Phi_h) \left({-\eta \over 2}\right) r^2 + {1 \over 6}\partial^2 U(\Phi_h) \left({-\eta \over 2}\right)^2 r^3 + \cdots  \right), \label{fgeneral}
\end{align}
for some constant $w_0$.

We are interested in solutions where $r = 0$ corresponds to a horizon, which we can arrange for by taking $w_0 \rightarrow - \infty$. Let's study the thermodynamics of this general model. Taking $t$ to be the correct asymptotic time, the temperature of the solution is
\begin{align}
T = {1 \over 4 \pi } \partial_r f(r)|_{r \rightarrow 0} = { |U(\Phi_h)| \over 2  \pi \eta }.
\end{align}
Thus, we see that the zero temperature solution corresponds to $U(\Phi_h) = 0$; indeed, when this happens, $f(r)$ has a double zero at the horizon. We label this value of the dilaton as $\Phi_0$. We can perform another expansion around the zero temperature solution, $\Phi^2_h = \Phi^2_0 + \delta \Phi^2$, so that 
\begin{align}
U(\Phi_h) = \partial U(\Phi_0) \delta \Phi^2 + \cdots,
\end{align}
or $\delta \Phi^2 = 2 \pi \eta  T/|\partial U(\Phi_0)|$. Using the Wald formula to compute the entropy, one finds
\begin{align}
S = {\Phi_h^2 \over 4 G_2} = {1 \over 4 G_2} \left( \Phi^2_0 + {2 \pi \eta  T \over |\partial U(\Phi_0)|}  + ...\right),
\end{align}
for small temperatures $T$. This shows that it is a general result that the entropy of near-extremal black holes has a linear dependence on  $T$ at low temperatures. Working in the canonical ensemble, one can use the first law of thermodynamics to find the energy above extremality to be
\begin{align}
\Delta E = {\pi \eta  \over 4 G_2 |\partial U(\Phi_0)|} T^2 + \cdots,
\end{align}
which gives 
\begin{align} \label{mgap_general}
M_{gap} = {4G_2|\partial U(\Phi_0)| \over \pi \eta}.
\end{align}
Note that the low-temperature thermodynamic properties of near-extremal black holes above have been determined purely by near-horizon data, up to an ambiguity of $E(0)$, the mass of the black hole at zero temperature. This ambiguity is expected, because the mass of the black hole is determined not just by near-horizon data but by near-boundary data and the full action including boundary counterterms. Nonetheless, we see that $\Delta E$ and in particular $M_{gap}$ are determined solely by near-horizon data and therefore by the JT model.

We can also compute the conformal symmetry breaking scale of this general model. Setting $\lambda = f(\Phi) = 0$ and linearizing the action around the zero temperature solution, we find the following action
\begin{align}
S = {1 \over 16 \pi G_2} \int dt dr  \left( - {U(\Phi) \over 2} h^2  + (\Phi^2)' e^{2 w} g \partial_r h  \right),
\end{align} 
which, after integrating out $h$, becomes
\begin{align}
S = {1 \over 16 \pi G_2} \int dt dr {\left[(\Phi^2)' e^{2 w}\right]^2 \over 2 U(\Phi)} (g')^2.
\end{align}
Near the horizon, we can expand this to obtain
\begin{align}
S &= {1 \over 16 \pi G_2} \int dt dr\left({\eta \left[- \partial U(\Phi_0)\right] \over 16}r^3 (g')^2  + \mO(r^4) \right)\\
&= {1 \over 16 \pi G_2} \int dt dz   \left( {\eta  \over 4 \left[- \partial U(\Phi_0)\right] z } (\partial_z g)^2 + \mO({1 \over z^2}) \right)
\end{align}
where we transformed to the Poincare coordinates defined by $z =  \left(  2 \over - \partial U(\Phi_0) \right) {1 \over r}$ in the second line. We can read off the breaking scale to be
\begin{align}
E_{br} \sim {G_2|\partial U(\Phi_0)| \over \eta}
\end{align}
in agreement with the mass gap \eqref{mgap_general}. This constitutes a proof of the equality of the mass gap and breaking scale for all models whose IR physics is governed by the JT model.\footnote{As mentioned in section \ref{comparison}, an important implicit assumption here is that the matter propagator of the full geometry approaches that of the constant dilaton $AdS_2$ in the IR. We believe that this is true at least at tree level.}

Finally, we show how to identify the parameters in the examples of section \ref{comparison} to those of the JT model in section \ref{APreview}. First, note that by the following coordinate transformation
\be
z = -{1\over 2}\sqrt{a \over 2\mu} \ln\left({\tilde{r} \over \tilde{r}+(4/C)\sqrt{2\mu/a}}\right),
\ee
\eqref{finite-temp metric} and \eqref{finite-temp dilaton} can be put into the form
\begin{align}
ds^2 &= -{\tilde{r}(\tilde{r}+(4/C)\sqrt{2\mu / a}) \over 2/C} dt^2 + {2/C \over \tilde{r}(\tilde{r}+(4/C)\sqrt{2\mu / a})}d\tilde{r}^2, \label{finite-temp metric2} \\ 
\Phi^2 &= \Phi_0^2 + \sqrt{2\mu a} +{aC \over 2} \tilde{r}. \label{finite-temp dilaton2}  
\end{align}

On the other hand, a general near-extremal black hole metric in $n+1$ dimensions can be written as
\be \label{higher-ansatz}
ds_{n+1}^2 = -V(r)dt^2 +{dr^2 \over V(r)} + \phi(r)^2dx_{n-1}^2,  
\ee
where
\be
V(r) = {(r-r_+)(r-r_-) \over L(r)^2}.
\ee
$V(r)$ has two real roots $r_{\pm}$ and $L(r)$ is a smooth function nonvanishing at $r=r_+$. $dx_{n-1}^2$ is a $(n-1)$-dimensional metric having a dimensionless volume $w_{n-1}$. As in \eqref{btz-reduction}, one can also consider a non-diagonal reduction where $(d\theta+A_{\mu}dx^{\mu})^2$ for an internal direction $\theta$ replaces $dx_{n-1}^2$. What follows is unchanged in this case with $w_{n-1}$ being the coordinate volume of $S^1$ parametrized by $\theta$.

The dimensional reduction of the higher dimensional theory with the ansatz \eqref{higher-ansatz} takes the form \eqref{realistic} with 
\begin{align}
ds^2 &= -V(r)dt^2 + {dr^2 \over V(r)},\\
\Phi^2 &= {G_2 \over G_{n+1}} w_{n-1}\phi^{n-1},
\end{align} 
where $G_{n+1}$ is the higher dimensional Newton's constant. We specialize to $\phi(r) = r$ as in our examples. Before comparing with \eqref{finite-temp metric2}, we need to transform to the correct conformal gauge determined by removing the dilaton kinetic term using \eqref{Weyl_trans}. Thus, the metric becomes $ds_{new}^2 = \Phi^{\lambda/2}ds^2$.

The next step is to compare the near-horizon expansion of the near-extremal black hole solution with the finite-temperature solution of the JT model \eqref{finite-temp metric2}, \eqref{finite-temp dilaton2}. The key point to keep in mind when expanding the near-extremal black hole solution around the horizon is that $r_{+-} = r_+-r_-$ is also taken to be as small as $\Delta r = r-r_+$. Alternatively, one can think of this as a double expansion in $\Delta r$ and $r_{+-}$. Also, the fact that the black hole is in the canonical ensemble (i.e. fixed charge rather than fixed chemical potential) implies a constraint that $r_+r_- = r_0^2$, where $r_0$ is the horizon radius of the extremal black hole with the given charge. For the near-extremal black hole, this says $r_{+-} \simeq 2(r_+ - r_0)$. Taking these into account, the near-horizon expansion of the near-extremal black hole is given by
\begin{align}
ds_{new}^2 & = -{\tilde{r}(\tilde{r} + \tilde{r}_{+-}) \over \tilde{L}^2}dt^2 + {\tilde{L}^2 \over \tilde{r}(\tilde{r} + \tilde{r}_{+-})}d\tilde{r}^2,\label{near-extremal metric}\\
\Phi^2 &= \Phi_0^2 + {G_2 \over G_{n+1}\Phi_0^{\lambda/2}}w_{n-1}(n-1)r_0^{n-2}\left({\tilde{r}_{+-} \over 2} + \tilde{r}\right)\label{near-extremal dilaton},
\end{align}   
where $\Phi_0 = \Phi(r_0)$, $\tilde{L}^2 = \Phi_0^{\lambda/2}L(r_0)^2$, $\tilde{r} = \Phi_0^{\lambda/2}r$, and $\tilde{r}_{+-} = \Phi_0^{\lambda/2}r_{+-}$.

Comparing \eqref{finite-temp metric2}, \eqref{finite-temp dilaton2} with \eqref{near-extremal metric}, \eqref{near-extremal dilaton}, we find that
\begin{align}
{a \over G_2} &= {  w_{n-1}(n-1)L^2r_0^{n-2} \over G_{n+1}},\\
C &= {2 \over \Phi_0^{\lambda/2}L^2},\\
{\mu \over G_2} &= {1\over 8}{ w_{n-1}(n-1)r_0^{n-2}r_{+-}^2 \over G_{n+1} L^2}.
\end{align}
With this identification, \eqref{APgraviton} and \eqref{APthermo} indeed agree exactly with the graviton kinetic term and $\Delta E(T)$ in all the higher dimensional (near-)extremal black holes we considered in section \ref{comparison}, including the numerical coefficient.

\section{Summary and Discussion} \label{discussion}

In this paper, we have argued that the thermodynamic mass gap of near-extremal black holes and the breaking scale of the near-horizon $AdS_2$ conformal symmetry are in fact the same. The origins of these two scales are a priori rather different; the former is obtained from the black hole thermodynamics, while the latter is obtained from computing the connected four point function of a matter field at zero temperature. However, they are both intimately connected to the strong backreaction effect in $AdS_2$. 

Recall that Hawking's semiclassical calculation must break down at the mass gap despite the macroscopic size of the horizon since the remaining energy for the black hole at that point is smaller than the energy of the typical Hawking quantum. What goes wrong with Hawking's calculation in this case is the assumption that the black hole evaporation is sufficiently slow that the quantum field theory on a fixed background is a good approximation. At temperatures as low as the mass gap, the change in the black hole geometry due to outgoing Hawking radiation is not adiabatic, and its backreaction on the near-horizon $AdS_2$ throat is important.

On the other hand, the connected four point function breaks the apparent conformal symmetry of $AdS_2$ due to the matter field's backreaction on the metric. More precisely, the explicit breaking of the conformal symmetry in the UV due to the dilaton does not quite decouple from the IR physics. In the constant dilaton limit, where the conformal symmetry is restored, the connected four point function diverges, and is a manifestation of the fact that this four point function is sensitive to the strong backreaction in $AdS_2$.

Another hint that the two scales should coincide comes from the observation that the relation $E(T) = M_{gap}^{-1} T^2$ is reminiscent of 2d CFT \cite{Maldacena:1997ih}. In 2d CFT, $M_{gap}^{-1} = \pi cL/12$, where $L$ is the size of the system and $c$ is the central charge, and the thermal wavelength $1/T$ has to be shorter than the effective size of the system in order for that relation to be valid. In CFTs dual to black holes, the effective size is expected to be given not by the actual size $L$ but by $cL$, due to twisted sectors \cite{Maldacena:1996ds}. In fact, otherwise, the CFT would not be able to reproduce the thermodynamics of near-extremal black holes. On the other hand, the conformal symmetry of the CFT would also be broken at the scale of this effective size\footnote{That is, the underlying theory itself is still conformal, but its correlators will deviate from the conformal form of correlators on the plane.}. Now, if one boldly makes an analogy between the holographic $CFT_1$ arising from the near-horizon $AdS_2$ and the (chiral half of) 2d CFT, one is led to a conclusion that $M_{gap}$ and the conformal breaking scale in $AdS_2/CFT_1$ have to be the same scale\footnote{In fact, on the gravity side, one can study how the holographic stress tensor in the JT model transforms under the boundary local conformal transformation from which the central charge can be read off. Comparing this with $M_{gap}$ could provide evidence for or against the conjecture that $AdS_2$ is dual to a chiral half of 2d CFT \cite{futureAK}. This comparison is also equivalent to that of Cardy's formula and the Bekenstein-Hawking entropy.}. Although this analogy seems to fit well with other observations on $AdS_2/CFT_1$ \cite{Strominger:1998yg}, there are many questions that remain to be answered. For example, for near-extremal black holes whose near-horizon geometry does not contain $AdS_3$, it is less clear how obtain a $CFT_2$ from which the $CFT_1$ emerges. Another interesting question is why the holographic $CFT_1$ obeys Cardy's formula. Cardy's formula seems to reproduce the Bekenstein-Hawking entropy of black holes in $AdS_2$, despite the lack of a microscopic explanation \cite{Cadoni:1999ja,  Grumiller:2015vaa, Castro:2008ms}. This can be viewed as evidence that the holographic $CFT_1$ is closely related to 2d CFT, but we are still lacking an explanation. A better understanding of this phenomenon is likely to shed more light on $AdS_2/CFT_1$ as well as how it is related to 2d CFT. 

Aside from the statement that the mass gap is equal to the breaking scale, we also established that the JT model provides a universal description of near-horizon physics of near-extremal black holes. An important exception is Kerr black holes. Since some of their metric components depend nontrivially on compact directions, it is difficult to dimensionally reduce them to two dimensions\footnote{\cite{Castro:2009jf} considered a certain dimensional reduction of the four dimensional Einstein gravity, but their two dimensional theory does not seem to contain the full Kerr geometry but only its near-horizon geometry.}. However, despite the technical difficulty, we believe that the four point function in the (near-)extremal Kerr background can in principle be computed and will break the IR conformal symmetry. Moreover, we expect that the scale at which the IR conformal symmetry is broken will coincide with the mass gap of near-extremal Kerr black holes. It would be interesting to consider a generalization of the JT model that describes near-extremal Kerr black holes as well. The Kerr/CFT correspondence \cite{Guica:2008mu} in its current form is analogous to $AdS_2/CFT_1$ with constant dilaton. Indeed, \cite{Amsel:2009ev} shows that there is no finite energy state with NHEK(Near-Horizon Extreme Kerr) asymptotics. To control the backreaction, it seems necessary to allow the transverse directions to expand toward the boundary as in the JT model.

Throughout this paper, our discussion has largely been classical. One can use the Feynman approach developed in section \ref{BQFT} is compute the quantum corrections to various quantities. We saw in section \ref{universality} that the JT model correctly captures the classical IR physics of near-extremal black holes, and thus would be interesting to see how much of quantum corrections carry over. Due to the simplicity of the JT model, we can actually make some statements about the quantum corrections without having to explictly perform any calculations. Take for example the two point function of the operator $\mO$ dual to the scalar field and consider its quantum corrections coming from graviton/scalar loops. Since the graviton propagator is proportional to the scale $G/a$ the general form of the two point function will be
\begin{align}
\lan \mO(t) \mO(0) \ran  = {1 \over t^2}  \sum_{n = 0}^\infty a_n \left({G t \over a}\right)^n,
\end{align}
where $n$ can be viewed as the number of loops in the diagram. There's also the possibility of renormalization introducing terms logarithmic in $G t / a$. Notice that the ${G t / a}$ acts as a coupling constant of the theory. Therefore, the theory becomes strongly coupled in the IR once $t \sim a/G$.

It would also be particularly interesting to compute quantum corrections to the partition function $Z(\beta)$ at a finite temperature and see their effect on the thermodynamics. Again, we can make interesting statements simply from dimensional analysis. From dimensional analysis, again we have
\begin{align}\label{logZ}
\ln Z(\beta) &= -\beta F_0 + \ln Z_{loops}(\beta) \\
&= - \beta F_0+ \sum_{n = 0}^{\infty} c_n \left( {G \beta \over a} \right)^n
\end{align}    
where $c_n$ are dimensionless coefficients.\footnote{$S = d(T\ln Z)/dT$ is a shannon entropy which is always non-negative for finite systems. This implies that the exact non-perturbative free energy should give a non-negative entropy, although term by term in perturbation theory $S$ seems to diverge as $T$ goes to zero.} Where $F_0(\beta)$ is the free energy of the classical solution we have been studying in the previous sections. The $n = 0$ term comes from the one-loop determinant of the quantum fields and will be logarithmic in $G \beta / a$. The higher order terms will come from higher order loop diagrams. Again, from the form of the expansion \eqref{logZ}, we see that $G \beta /a$ acts like a coupling constant and the perturbation theory breaks down once $\beta \sim a/G$ or when the temperature reaches the mass gap. In particular, precisely at the mass gap, the quantum fluctuations start to dominate over the classical contributions. This seems to be another indication that the near-horizon region of a near-extremal black hole below the mass gap is no longer semiclassical even when the horizon is macroscopic. This means that the semiclassical picture of the (near-horizon region of) extremal black holes might be misleading, although we used it throughout this paper. Perhaps, the ideas along the line of the fuzzball proposal (see, for example, \cite{Mathur:2009hf} and references therein) might provide a more accurate picture of them in certain contexts.  

\acknowledgments
We are grateful to Xi Dong, Blaise Gout\'eraux, Guy Gur-Ari, Sean Hartnoll, Hong Liu, Juan Maldacena, Don Marolf, Joe Polchinski, Steve Shenker,  Eva Silverstein, and Douglas Stanford for useful discussions. B.K. is supported by the National Science Foundation under grant PHY-0756174 and NSF PHY11-25915. 	

\appendix
\section{No transverse gauge perturbation in spherical or toroidally symmetric ansatz} \label{appendixA}
Consider the most general perturbation around a spherically or toroidally symmetric Einstein-Maxwell background in $3+1$ dimensions that respect the symmetry. In $3+1$ dimensions, non-vanishing transverse components of the field strength (i.e. transverse to $(t,r)$ directions), or magnetic fields, can be consistent with the symmetry. We argue that transverse gauge perturbations can be consistently set to zero. Our argument below works for both spherical and toroidal symmetry, but for definiteness we consider the toroidally symmetric case.   

One set of Maxwell's equation,
\begin{align}
\nabla_{[\mu} F_{\nu\sigma]}=0,
\end{align}
can be simplified using (anti)symmetry properties of Christoffel symbols and $F_{\mu\nu}$ to
\begin{align}
\partial_{\mu}F_{\nu\sigma} + \partial_{\nu}F_{\sigma\mu} + \partial_{\sigma}F_{\mu\nu} = 0.
\end{align}
By assumption of the toroidal symmetry, only $F_{tr}$ and $F_{xy}$ can be nonzero and they depend only on $(t,r)$. Then, by choosing $(\mu,\nu,\sigma)$ to be $(t,x,y)$ and $(r,x,y)$, we see that
\begin{align}
&\partial_tF_{xy}=0,\\
&\partial_rF_{xy}=0.
\end{align}
This means that $F_{xy}$ is constant everywhere in the spacetime. Therefore, we can consistently set the transverse perturbation of $F_{xy}$ to zero.

\section{Classical equivalence of dilaton gravity theories with and without gauge fields} \label{appendixB}
Consider\footnote{If there is a dilaton kinetic term, one can eliminate it by a Weyl tranformation to get this form.}
\be \label{Maxwell-dilaton}
S = \int d^2x \sqrt{-g}[\Phi^2 R - U(\Phi) - f(\Phi)F^2]. 
\ee
At classical level, we can solve the gauge equations of motion (EOM), and plug back the solution into the other EOM for the dilaton and metric. We show that, as a result of this, the above dilaton gravity theory is equivalent at the level of EOM to another one without gauge fields. This conclusion also holds when we add neutral matter fields. In particular, Hartman-Strominger (HS) model \cite{Hartman:2008dq}, which corresponds to $U(\Phi) = -C\Phi^2$ and $f(\Phi)=1$, is classically equivalent to the JT model, corresponding to $U(\Phi)=A-C\Phi^2$ and $f(\Phi)=0$.
The variations of $S$ with respect to metric, gauge field, and dilaton are as follows:
\begin{align}
{\delta S \over \delta g^{\mu \nu}} &= \sqrt{-g} \left[- \nabla_{\mu} \nabla_{\nu} \Phi^2 + \nabla^2 \Phi^2 g_{\mu \nu}  + {1 \over 2} U(\Phi)  g_{\mu \nu} 
 \right.  \nonumber\\
& \ \ \left. +f(\Phi)\left({1\over 2} F^2g_{\mu \nu}  - 2F_{\mu \sigma} F_{\nu}^{\ \sigma}\right) \right],\\
{\delta S \over \delta \Phi} & = \sqrt{-g} [2\Phi R - U'(\Phi) - f'(\Phi)F^2],\\
{\delta S \over \delta A_{\nu}} &= 4\partial_{\mu} (\sqrt{-g}f(\Phi)F^{\mu \nu}).
\end{align}
Solving the gauge EOM gives
\be
f(\Phi) F^{\mu \nu} = E \epsilon ^{\mu \nu},
\ee
for some constant $E$. Plugging this solution in the metric and dilaton EOM gives
\begin{align}
{\delta S \over \delta g^{\mu \nu}} &= \sqrt{-g} \left[- \nabla_{\mu} \nabla_{\nu} \Phi^2 + \nabla^2 \Phi^2 g_{\mu \nu}  + {1 \over 2} U(\Phi)  g_{\mu \nu} 
 \right.  \nonumber\\
& \ \ \left. +f(\Phi)^{-1}E^2 g_{\mu \nu}  \right],\\
{\delta S \over \delta \Phi} & = \sqrt{-g} [2\Phi R - U'(\Phi) JT model+ 2f'(\Phi)f(\Phi)^{-2}E^2].
\end{align}
By comparing the two sets of EOM before and after solving the gauge EOM, we conclude that the dilaton gravity theory with $U(\Phi)$ and $f(\Phi) \neq 0$ is classically equivalent to another one with $U(\Phi) \rightarrow U(\Phi) + 2E^2/f(\Phi)$ and $f(\Phi) \rightarrow 0$. In particular, this implies that HS model is classically equivalent to the JT model with $A=2E^2$.

\section{Further comments about the thermodynamic mass gap} \label{appendixC}
When considering the thermodynamic mass gap, one has to be careful about whether the system is in the canonical or grandcanonical ensemble. Throughout this paper, we considered the canonical ensemble. In this case, since $dE = TdS$, $S(T) - S(0)$ is determined by $E(T)$ and gives constraint $\alpha > 1$ when $E(T) \sim T^{\alpha}$ at low temperatures. On the other hand, in the grandcanonical ensemble, $dE = TdS+\Phi dQ$, so $\alpha$ does not have to be necessarily greater than one. For example, in case of charged spherical or planar black holes in $AdS$ at a fixed chemical potential, $E(T) \sim T$ at low temperatures. 	    

Another important caveat in this discussion is whether the black hole geometry is the dominant saddle \cite{Chamblin:1999tk}. In the canonical ensemble, near-extremal black holes are the dominant saddle at sufficiently low temperatures at any nonzero charge. In the grandcanonical ensemble, however, the same is true only above a certain critical value of the chemical potential. Below this critical value, there exists an analog of Hawking-Page transition such that the dominant saddle at sufficiently low temperatures is the pure $AdS$. In the latter case, there is no black hole in the bulk, and thus the argument against the validity of Hawking's semiclassical calculation below the mass gap simply no longer applies.  Moreover, since the temperature-dependent part of $E(T)$ coming from the bulk fields will have coefficients of order one, the mass gap deduced from it will not be interesting in the sense that it is not parametrically smaller than the inverse size of the boundary system.

So far, we have discussed asymptotically $AdS$ black holes in which case we were able to explicitly see that $\alpha$ is an integer and equal to $2$ at low temperatures.  One may ask whether there exist holographic systems dual to near-extremal black holes at low temperatures that have non-integer values of $\alpha$. Especially interesting in this regard are Lifshitz black holes with or without hyperscaling violation \cite{Taylor:2015glc}. 
The Schwarzschild analog of these black holes (i.e. those whose blackening factor has a single simple real root) exhibits a scaling relation for the entropy as a function of temperature, $S(T) \sim T^{(d-\theta)/z}$, where $d$ is the number of the spatial dimensions in the boundary theory, $z$ the dynamical exponent, and $\theta$ the hyperscaling exponent. Depending on the values of $d, z, \theta$, one can have arbitrary scaling relations, up to the constraint on these parameters from physical conditions like null energy condition and thermodynamic stability. However, at low temperatures, these black holes have a small horizon area and the spacetime near the horizon is too strongly coupled to trust the semiclassical approximation. In fact, one has to be first careful about whether the black hole spacetime is the dominant saddle at low temperatures.

To have a meaningful discussion of the thermodynamic mass gap, one has to look at the Reissner-Nordstrom analog of these black holes, which may have a macroscopic horizon at low temperatures \cite{Tarrio:2011de, Alishahiha:2012qu}. Depending on $z$ and $\theta$, the qualitative features of the phase diagram in the canonical and grandcanonical ensembles may differ significantly from the $AdS$ case. But, regardless of the structure of the phase diagram, we can make the following general comments. If a non-black-hole spacetime is the dominant saddle at low temperatures, the whole mass gap issue is trivial for the same reasons as above. If the black hole spacetime is the dominant saddle and it has a macroscopic horizon in the zero-temperature limit, its near-horizon geometry is $AdS_2$ times some internal manifold, $S(T) \sim T$, and in case of the canonical ensemble $E(T) \sim T^2$ as well. Note that the behavior of $S(T)$ and $E(T)$ at low temperatures is identical to the asymptotically $AdS$ case even when $z$ and $\theta$ are nontrivial. This is because the near-horizon geometry of the near-extremal black holes contains $AdS_2$. In fact, the relation $S(T) \sim T$ at low temperatures follows very generally from the assumption that the horizon area is nonvanishing at zero temperature. Let $r$ be the horizon radius. Since the entropy is given by the horizon area,
\be
S(T) - S(0) \sim V(r^n - r_0^n) \sim Vn r_0^{n-1}\Delta r,
\ee 
where $V$ is the coordinate volume of the transverse manifold, $n$ the number of the transverse directions, $r_0$ the horizon radius at zero temperature, and $\Delta r = r-r_0$. On the other hand, $T \sim (dT/dr)_{r=r_0}\Delta r$, where $(dT/dr)_{r=r_0}$ is not zero since $T$ as a function of $r$ always has a simple zero at $r=r_0$. See section \ref{universality} for a more general treatment. Therefore, we have $S(T) \sim T$ at low temperatures.

\bibliographystyle{jhep}
\bibliography{bibliography}

\providecommand{\href}[2]{#2}\begingroup\raggedright\begin{thebibliography}{10}

\bibitem{Strominger:1996sh}
A.~Strominger and C.~Vafa, {\it {Microscopic origin of the Bekenstein-Hawking
  entropy}},  {\em Phys. Lett.} {\bf B379} (1996) 99--104,
  [\href{http://arxiv.org/abs/hep-th/9601029}{{\tt hep-th/9601029}}].

\bibitem{Maldacena:1997re}
J.~M. Maldacena, {\it {The Large N limit of superconformal field theories and
  supergravity}},  {\em Int. J. Theor. Phys.} {\bf 38} (1999) 1113--1133,
  [\href{http://arxiv.org/abs/hep-th/9711200}{{\tt hep-th/9711200}}]. [Adv.
  Theor. Math. Phys.2,231(1998)].

\bibitem{Jensen:2011su}
K.~Jensen, S.~Kachru, A.~Karch, J.~Polchinski, and E.~Silverstein, {\it
  {Towards a holographic marginal Fermi liquid}},  {\em Phys. Rev.} {\bf D84}
  (2011) 126002, [\href{http://arxiv.org/abs/1105.1772}{{\tt
  arXiv:1105.1772}}].

\bibitem{Maldacena:1998uz}
J.~M. Maldacena, J.~Michelson, and A.~Strominger, {\it {Anti-de Sitter
  fragmentation}},  {\em JHEP} {\bf 02} (1999) 011,
  [\href{http://arxiv.org/abs/hep-th/9812073}{{\tt hep-th/9812073}}].

\bibitem{Almheiri:2014cka}
A.~Almheiri and J.~Polchinski, {\it {Models of AdS$_{2}$ backreaction and
  holography}},  {\em JHEP} {\bf 11} (2015) 014,
  [\href{http://arxiv.org/abs/1402.6334}{{\tt arXiv:1402.6334}}].

\bibitem{Jackiw:1984je}
R.~Jackiw, {\it {Lower Dimensional Gravity}},  {\em Nucl. Phys.} {\bf B252}
  (1985) 343--356.

\bibitem{Teitelboim:1983ux}
C.~Teitelboim, {\it {Gravitation and Hamiltonian Structure in Two Space-Time
  Dimensions}},  {\em Phys. Lett.} {\bf B126} (1983) 41--45.

\bibitem{Preskill:1991tb}
J.~Preskill, P.~Schwarz, A.~D. Shapere, S.~Trivedi, and F.~Wilczek, {\it
  {Limitations on the statistical description of black holes}},  {\em Mod.
  Phys. Lett.} {\bf A6} (1991) 2353--2362.

\bibitem{Maldacena:1996ds}
J.~M. Maldacena and L.~Susskind, {\it {D-branes and fat black holes}},  {\em
  Nucl. Phys.} {\bf B475} (1996) 679--690,
  [\href{http://arxiv.org/abs/hep-th/9604042}{{\tt hep-th/9604042}}].

\bibitem{Maldacena:1997ih}
J.~M. Maldacena and A.~Strominger, {\it {Universal low-energy dynamics for
  rotating black holes}},  {\em Phys. Rev.} {\bf D56} (1997) 4975--4983,
  [\href{http://arxiv.org/abs/hep-th/9702015}{{\tt hep-th/9702015}}].

\bibitem{Jensen:2016pah}
K.~Jensen, {\it {Chaos and hydrodynamics near AdS$_2$}},
  \href{http://arxiv.org/abs/1605.0609}{{\tt arXiv:1605.0609}}.

\bibitem{Maldacena:2016upp}
J.~Maldacena, D.~Stanford, and Z.~Yang, {\it {Conformal symmetry and its
  breaking in two dimensional Nearly Anti-de-Sitter space}},
  \href{http://arxiv.org/abs/1606.0185}{{\tt arXiv:1606.0185}}.

\bibitem{Engelsoy:2016xyb}
J.~Engelsöy, T.~G. Mertens, and H.~Verlinde, {\it {An Investigation of AdS2
  Backreaction and Holography}},  \href{http://arxiv.org/abs/1606.0343}{{\tt
  arXiv:1606.0343}}.

\bibitem{Page:2000dk}
D.~N. Page, {\it {Thermodynamics of near extreme black holes}},  2000.
\newblock \href{http://arxiv.org/abs/hep-th/0012020}{{\tt hep-th/0012020}}.

\bibitem{'tHooft:1993gx}
G.~'t~Hooft, {\it {Dimensional reduction in quantum gravity}},  in {\em
  {Salamfest 1993:0284-296}}, pp.~0284--296, 1993.
\newblock \href{http://arxiv.org/abs/gr-qc/9310026}{{\tt gr-qc/9310026}}.

\bibitem{Susskind:1994vu}
L.~Susskind, {\it {The World as a hologram}},  {\em J. Math. Phys.} {\bf 36}
  (1995) 6377--6396, [\href{http://arxiv.org/abs/hep-th/9409089}{{\tt
  hep-th/9409089}}].

\bibitem{Banados:1992wn}
M.~Banados, C.~Teitelboim, and J.~Zanelli, {\it {The Black hole in
  three-dimensional space-time}},  {\em Phys. Rev. Lett.} {\bf 69} (1992)
  1849--1851, [\href{http://arxiv.org/abs/hep-th/9204099}{{\tt
  hep-th/9204099}}].

\bibitem{Strominger:1998yg}
A.~Strominger, {\it {AdS(2) quantum gravity and string theory}},  {\em JHEP}
  {\bf 01} (1999) 007, [\href{http://arxiv.org/abs/hep-th/9809027}{{\tt
  hep-th/9809027}}].

\bibitem{Castro:2008ms}
A.~Castro, D.~Grumiller, F.~Larsen, and R.~McNees, {\it {Holographic
  Description of AdS(2) Black Holes}},  {\em JHEP} {\bf 11} (2008) 052,
  [\href{http://arxiv.org/abs/0809.4264}{{\tt arXiv:0809.4264}}].

\bibitem{KeskiVakkuri:1998nw}
E.~Keski-Vakkuri, {\it {Bulk and boundary dynamics in BTZ black holes}},  {\em
  Phys. Rev.} {\bf D59} (1999) 104001,
  [\href{http://arxiv.org/abs/hep-th/9808037}{{\tt hep-th/9808037}}].

\bibitem{Chamblin:1999tk}
A.~Chamblin, R.~Emparan, C.~V. Johnson, and R.~C. Myers, {\it {Charged AdS
  black holes and catastrophic holography}},  {\em Phys. Rev.} {\bf D60} (1999)
  064018, [\href{http://arxiv.org/abs/hep-th/9902170}{{\tt hep-th/9902170}}].

\bibitem{Faulkner:2009wj}
T.~Faulkner, H.~Liu, J.~McGreevy, and D.~Vegh, {\it {Emergent quantum
  criticality, Fermi surfaces, and AdS(2)}},  {\em Phys. Rev.} {\bf D83} (2011)
  125002, [\href{http://arxiv.org/abs/0907.2694}{{\tt arXiv:0907.2694}}].

\bibitem{futureAK}
A.~Almheiri and B.~Kang, {\it {work in progress}}, .

\bibitem{Cadoni:1999ja}
M.~Cadoni and S.~Mignemi, {\it {Asymptotic symmetries of AdS(2) and conformal
  group in d = 1}},  {\em Nucl. Phys.} {\bf B557} (1999) 165--180,
  [\href{http://arxiv.org/abs/hep-th/9902040}{{\tt hep-th/9902040}}].

\bibitem{Grumiller:2015vaa}
D.~Grumiller, J.~Salzer, and D.~Vassilevich, {\it {AdS$_{2}$ holography is
  (non-)trivial for (non-)constant dilaton}},  {\em JHEP} {\bf 12} (2015) 015,
  [\href{http://arxiv.org/abs/1509.0848}{{\tt arXiv:1509.0848}}].

\bibitem{Castro:2009jf}
A.~Castro and F.~Larsen, {\it {Near Extremal Kerr Entropy from AdS(2) Quantum
  Gravity}},  {\em JHEP} {\bf 12} (2009) 037,
  [\href{http://arxiv.org/abs/0908.1121}{{\tt arXiv:0908.1121}}].

\bibitem{Guica:2008mu}
M.~Guica, T.~Hartman, W.~Song, and A.~Strominger, {\it {The Kerr/CFT
  Correspondence}},  {\em Phys. Rev.} {\bf D80} (2009) 124008,
  [\href{http://arxiv.org/abs/0809.4266}{{\tt arXiv:0809.4266}}].

\bibitem{Amsel:2009ev}
A.~J. Amsel, G.~T. Horowitz, D.~Marolf, and M.~M. Roberts, {\it {No Dynamics in
  the Extremal Kerr Throat}},  {\em JHEP} {\bf 09} (2009) 044,
  [\href{http://arxiv.org/abs/0906.2376}{{\tt arXiv:0906.2376}}].

\bibitem{Mathur:2009hf}
S.~D. Mathur, {\it {The Information paradox: A Pedagogical introduction}},
  {\em Class. Quant. Grav.} {\bf 26} (2009) 224001,
  [\href{http://arxiv.org/abs/0909.1038}{{\tt arXiv:0909.1038}}].

\bibitem{Hartman:2008dq}
T.~Hartman and A.~Strominger, {\it {Central Charge for AdS(2) Quantum
  Gravity}},  {\em JHEP} {\bf 04} (2009) 026,
  [\href{http://arxiv.org/abs/0803.3621}{{\tt arXiv:0803.3621}}].

\bibitem{Taylor:2015glc}
M.~Taylor, {\it {Lifshitz holography}},  {\em Class. Quant. Grav.} {\bf 33}
  (2016), no.~3 033001, [\href{http://arxiv.org/abs/1512.0355}{{\tt
  arXiv:1512.0355}}].

\bibitem{Tarrio:2011de}
J.~Tarrio and S.~Vandoren, {\it {Black holes and black branes in Lifshitz
  spacetimes}},  {\em JHEP} {\bf 09} (2011) 017,
  [\href{http://arxiv.org/abs/1105.6335}{{\tt arXiv:1105.6335}}].

\bibitem{Alishahiha:2012qu}
M.~Alishahiha, E.~O~Colgain, and H.~Yavartanoo, {\it {Charged Black Branes with
  Hyperscaling Violating Factor}},  {\em JHEP} {\bf 11} (2012) 137,
  [\href{http://arxiv.org/abs/1209.3946}{{\tt arXiv:1209.3946}}].

\end{thebibliography}\endgroup

\end{document}